\newcommand\kms{$\rm\,km\,s^{-1}\,$}

\newcommand\etal{{\it et al.}\ }

\documentstyle[11pt,aaspp4]{article}

\slugcomment{(to appear in the Astronomical Journal)}

\lefthead{Scodeggio, Giovanelli \& Haynes}
\righthead{Universality of the Fundamental Plane. Spectroscopy}

\begin{document}

\title{The Universality of the Fundamental Plane of E and S0 Galaxies.\\
Spectroscopic Data\footnote{
Work based in part on observations obtained with the Hale 200'' telescope 
at the Palomar Observatory. These observations were made as part of a 
continuing collaborative agreement between the California Institute of 
Technology and Cornell University.} }

\author{Marco Scodeggio\altaffilmark{2}, Riccardo Giovanelli and Martha P. 
Haynes}
\affil{Center for Radiophysics and Space Research and National Astronomy
and Ionosphere Center\altaffilmark{3}, Cornell University, Ithaca, NY 14853\\
mscodegg@eso.org \\ riccardo, haynes@astrosun.tn.cornell.edu}

\altaffiltext{2}{Present address: European Southern Observatory, 
Karl-Schwarzschild-Stra$\beta$e 2, D-85748, Garching bei M{\"u}nchen, 
Germany.}
\altaffiltext{3}{The National Astronomy and Ionosphere Center is
operated by Cornell University under a cooperative agreement with the
National Science Foundation.}

\begin{abstract}
We present here central velocity dispersion measurements for 325
early-type galaxies in eight clusters and groups of galaxies,
including new observations for 212 galaxies. The clusters and groups
are the A262, A1367, Coma (A1656), A2634, Cancer and Pegasus clusters, and
the NGC 383 and NGC 507 groups. The new measurements were derived from
medium dispersion spectra, that cover 600 \AA\ centered on the Mg Ib
triplet at $\lambda \sim 5175$.  Velocity dispersions were measured
using the Tonry \& Davis cross-correlation method, with a typical
accuracy of 6\%. A detailed comparison with other data sources is
made.
\end{abstract}


\section{Introduction}

This is the fourth in a series of papers (Scodeggio, Giovanelli \&
Haynes 1997a,b\markcite{pap1}\markcite{pap2}; hereafter Paper I and
Paper II; Scodeggio, Giovanelli \& Haynes 1998a\markcite{pap3};
hereafter Paper III) devoted to a study of the universality of the
Fundamental Plane (FP) relation of early-type galaxies (Djorgovski \&
Davis 1987\markcite{DD}; Dressler \etal 1987\markcite{7s}).  The
universality of the FP, and of other distance indicator relations like
the Tully-Fisher (TF) relation (Tully \& Fisher 1977\markcite{TF}), is
the fundamental assumption behind their use, but this assumption
cannot be taken for granted, because of the existence of well known
environmental effects that can link the evolution of a galaxy to the
properties of its surrounding environment. It is therefore important
to obtain an observational verification of such universality
(discussions on this subject are presented by Djorgovski, de Carvalho
\& Han 1988\markcite{Djorg88}; Guzm{\'a}n \etal 1992\markcite{Guz92}; 
J{\o}rgensen, Franx \& Kj{\ae}rgaard 1996\markcite{Jor96}; Schroeder 
1996\markcite{Schro96}; Giovanelli \etal 1997b\markcite{G97b}; Paper I).

In this series, we present the results of a FP study of eight
clusters, with the aim of obtaining a quantitative verification of the
FP universality through a comparison of accurate distance estimates
obtained using both the FP and the TF relations. These clusters are
taken from the list of clusters studied by Giovanelli \etal
(1997a,b\markcite{G97a}\markcite{G97b}) using the TF relation, and are
the Coma, A1367, A2634, A262, Cancer, and Pegasus clusters, and the
NGC 507 and NGC 383 groups. In this paper we present spectroscopic
data, necessary for building the FP relation, for 325 galaxies,
including new observations for 212 galaxies, while Paper III contains
I band CCD photometric data, and describes the cluster selection and
membership criteria adopted for the FP work.  In successive work
(Scodeggio, Giovanelli \& Haynes 1998b\markcite{pap5}) we will discuss
the properties of the I band FP relation, and the comparison between
TF and FP distance estimates for the clusters in our sample.

This papers is organized as follows: sections 2, 3, and 4 present our
new spectroscopic observations and the adopted data reduction
procedures.  Sections 5 and 6 present the complete spectroscopic
sample that will be used in the FP work, and the internal and external
comparisons used to assess the quality of our new measurements.

\section{Observations}

New spectroscopic observations for this work were obtained during 8
observing runs with the Hale 5m telescope of the Palomar Observatory,
from September 1992 to September 1996. Table 1 summarizes these
observing runs, and gives the parameters of the relative instrumental
setups.  During all runs the red camera of the Double Spectrograph
(Oke \& Gunn 1982\markcite{dbsp}) was used with a 1200 lines/mm
grating, to produce spectra with a dispersion of approximately 0.82
\AA\ per pixel (except in runs 7 and 8, when a different CCD chip was
mounted on the camera, and the dispersion was approximately 0.66 \AA\
per pixel).  This resulted in a spectral coverage of approximately 600
\AA, from 5000 to 5600 \AA, including the Mg Ib lines at 
$\lambda \sim 5175$. The spectral resolution, measured both from the
width of the calibration lamps lines, and from the width of the peak
in the autocorrelation function of the spectrum of velocity standard
stars, was approximately 2.2 \AA\ (FWHM) for all runs. This is
equivalent to a velocity resolution of 129 \kms at 5300 \AA. All
observations were obtained with a 2 arcsec wide slit, while the length
of the slit was 128 arcsec. The slit orientation was kept in the
East-West direction, unless a close pair of galaxies could be observed
at once with the proper slit orientation. Galaxies were visually
centered in the slit using the acquisition camera of the Double
Spectrpgraph.  Integration times varied between 10 and 90 minutes,
depending on the brightness of the galaxy. The median integration time
was 30 minutes.  He-Ne-Ar lamp spectra were obtained before and after
each galaxy observation, to provide wavelength calibration.  Late G
and early K type giant stars were observed each night to serve as
templates for radial velocity and velocity dispersion
determinations. To maximize the signal-to-noise ratio in these
template spectra, the star observations were obtained with 30-120 sec
integrations, while trailing the star along the slit.  This allowed to
spread the light on a large area of the CCD chip without saturating
the image.

\section{Basic data reduction and wavelength calibration}

The data reduction process was performed entirely using standard
IRAF\footnote{IRAF (Image Reduction and Analysis Facility) is
distributed by the National Optical Astronomy Observatories, which are
operated by the Association of Universities for Research in Astronomy
(AURA), Inc., under a cooperative agreement with the National Science
Foundation.} procedures.  All frames were bias-subtracted and
flat-fielded, using dome flats.  The number of pixels affected by
cosmic rays hits was small in all the frames, and they were blanked
during a phase of visual inspection of all frames.  Because most of
the spectroscopic observations were obtained in non photometric
conditions, no attempt was made to obtain a flux calibration for any
of the derived spectra.

Wavelength calibration was performed in a two-step process. First, the
He-Ne-Ar lamp spectra frames obtained before and after each galaxy
spectrum were used to obtain a two-dimensional dispersion solution,
where a wavelength value is assigned to each pixel in the CCD frame.
Typically 12-14 spectral lines, spread uniformly over the entire
wavelength range, were used in the fit. The rms deviations between
fitted and true wavelength were of the order of 0.06--0.1 \AA\ in the
central portion of the frames (where the galaxy spectra are
positioned), and increased up to 0.25 \AA\ at the edges of the frame,
due to the camera astigmatism and to focusing problems caused by the
less than perfect flatness of the CCD chip. Single line deviations
from the best fit solutions were always less than 0.3 \AA. An
uncertainty in the wavelength calibration of 0.1 \AA\ at 5300 \AA\
translates into a velocity uncertainty of 5.6 \kms. The CCD chip used
for runs 7 and 8 does not have focusing problems, and therefore the
rms uncertainty in the calibration of the spectra obtained in those
runs was of the order of 0.08 \AA, uniformly across the entire
frame. The two-dimensional dispersion solutions were applied to the
galaxy spectral frames, so that all pixels in one row of the corrected
CCD frame were characterized by the same wavelength.  Finally the two
strong night-sky lines at 5460.7 \AA\ (Hg) and 5577.4 \AA\ (OI) were
used to check the absolute wavelength calibration.  The 1-$\sigma$
dispersion in the measured wavelengths of these lines was
approximately 0.08 \AA. A small fraction (2-3\%) of the frames showed
large deviations ($> 4 \sigma$) in the wavelength measured for the two
sky lines.  For these cases the original dispersion solution was
rigidly shifted to bring the two lines into agreement with their
expected wavelengths.

An independent check on the wavelength calibration was performed using
the velocity standard star spectra. Radial velocities were measured
repeatedly using all star spectra obtained in the different runs.
Using in turn one star as velocity standard and all others as unknown,
radial velocities were derived, to be compared with the known radial
velocity of each star. The rms deviations between the measured and the
known velocities were between 6.0 and 8.5 \kms, when spectra obtained
during a single run only were considered, and approximately 9 \kms
when spectra from all runs were considered. This result is in good
agreement with the calibration uncertainty derived from the sky lines
measurements. Therefore 9 \kms is an estimate of the overall 
accuracy we can achieve with this spectroscopic data set.

\section{Determination of the spectroscopic parameters}

During the last twenty years a number of methods have been developed
to perform velocity dispersion measurements as objectively as possible
(see for example Sargent \etal 1977\markcite{SSBS}; Tonry \& Davis 
1979\markcite{TD}; Franx, Illingworth \& Hechman 1989\markcite{Franx89}; 
Bender 1990\markcite{Ben90}).  These methods are all
based on the comparison between the spectrum of the galaxy whose
velocity dispersion is to be determined, and a fiducial spectral
template. This can be either the spectrum of an appropriate star, with
spectral lines unresolved at the spectral resolution being used, or a
high signal-to-noise spectrum of a galaxy with known velocity
dispersion.  There are small but in principle significant differences
among these methods, but in practice all methods seem to give results
in good relative agreement, when applied to spectra with high
signal-to-noise ratio, making the choice of one particular method a
non critical issue.

In this work, quantitative measurements on the spectra have been
obtained using the cross-correlation technique of Tonry \& Davis
(1979\markcite{TD}), implemented in the IRAF task {\it fxcor}. The
basic assumption behind this and other similar methods is that the
spectrum of an elliptical galaxy (and also of the bulge of a disk
galaxy) is well approximated by the spectrum of its most luminous
stars (K0--K1 giants), modified only by the effects of the stellar
motions inside the galaxy. Since these motions introduce just a
Doppler shift in the stellar spectra, the galaxy spectrum is given by
the convolution of the spectrum of a K giant star with the line of
sight stellar velocity distribution (LOSVD). Therefore the LOSVD can
be obtained with a deconvolution process from the galaxy spectrum and
a suitable stellar template.  Since cross-correlation and convolution
are two related operators, the Tonry \& Davis algorithm uses
cross-correlation as a computational tool to derive the LOSVD. The
procedure takes advantage of the fact that the Fourier transform of a
convolution, or of a cross-correlation, reduces to a product between
individual functions' transforms.

The one-dimensional galaxy spectra used for the radial velocity and
velocity dispersion measurements were extracted from the
two-dimensional wavelength calibrated frames using a 6 arcsec wide
window, centered on the peak of the galaxy continuum light. This,
combined with the 2 arcsec slit aperture, provides a final effective
aperture of 6 x 2 arcsec for all the galaxy observations. The stellar
templates were obtained averaging the signal over 20--30 pixels in the
spatial dimension at all wavelengths, to produce spectra with a signal
to noise ratio greater than 200. Before computing the Fourier
transforms and the cross-correlation function, all spectra were
continuum-subtracted, normalized, end-masked with a cosine bell
function, and re-binned to a logarithmic wavelength scale. The final
velocity dispersion measurements have been obtained by averaging the
determinations obtained with five different stellar templates, of
spectral types between G9 and K2. No significant dependence of the
measurements on the template star spectral type was observed.

\subsection{The Effect of the Noise}

Because of the limited signal-to-noise (S/N) ratio of the spectra
being analyzed, and also to facilitate comparisons with other studies,
we have constrained the deconvolution procedure by assuming the LOSVD
to be Gaussian, and therefore characterized by two parameters only: a
mean redshift z and a velocity dispersion $\sigma$. The reliability of
this method has been tested using simulations with synthetic
spectra. Stellar templates have been broadened with Gaussian profiles
of known width to simulate a large range of velocity dispersions, and
noise has been added to the resulting spectra, to reproduce the S/N
characteristic of the galaxy spectra. Then velocity dispersions have
been measured using the original stellar spectrum as a template.
These tests show that {\it fxcor} produces an overestimate of the
velocity dispersion, at the 4-5\% level, while no bias is present in
the measurement of radial velocities. The raw measurements have been
therefore corrected to remove this effect. To a first approximation,
this effect appears to be independent of the S/N ratio of the input
spectrum, but no simulation was performed using spectra with S/N ratio
lower than 20.

If the true LOSVD is approximately Gaussian, then the statistical
uncertainty with which the velocity dispersion is obtained is a
function of the S/N ratio of the galaxy spectrum only, since the S/N
ratio in the template spectrum is many times higher than that of the
galaxy spectrum (see Tonry \& Davis 1979\markcite{TD} for details). An
estimate of the amplitude of this uncertainty was obtained using
synthetic spectra with added noise. It was found that a good
approximation to the value of the uncertainty is given by the relation
$\epsilon_{\log \sigma}\simeq 1.1 / (S/N)$.  The median uncertainty on
the radial velocity determination is 13 \kms, and on the velocity
dispersion is 6\%.

As a visual guide to the quality of the data presented in this work,
we show in Figure 1 the spectra of 6 different galaxies. These are
divided into two groups, and all three galaxies in one group have very
similar velocity dispersion (approximately 140 and 230 \kms,
respectively), but their spectra have different S/N ratio. These
spectra were chosen to provide an example of a spectrum at the upper
quartile, median, and lower quartile in the distribution of S/N values
for the spectra in our dataset.

\subsection{The Effect of Rotation}

Galaxy one-dimensional spectra were extracted using a 6 arcsec
aperture, which can bracket a fairly large portion of the target
galaxy, especially for more distant clusters. Therefore galaxy
rotation, if present, can contribute some broadening of the spectral
lines and bias the velocity dispersion measurements.  We thus use the
galaxy two-dimensional spectra to estimate the galaxy rotation
velocity, and gauge whether a correction of the velocity dispersion
measurements is necessary. A 5 point rotation curve within the 6
arcsec aperture (10 pixels) has been obtained for each galaxy, using 5
one-dimensional spectra, extracted using 2-pixel wide apertures
positioned side by side. Using the cross-correlation method,
individual mean velocities can be obtained for the 5 spectra. The
differences in the radial velocities measured between each one of the
four lateral spectra and the one measured in the central 2 pixels
define the projection of the galaxy rotation curve along the slit.

The contribution of a given rotation to the broadening of the spectral
lines has been estimated using a very simple model. A galaxy spectrum
has been simulated combining 5 copies of a stellar template, broadened
with a Gaussian of fixed width, and shifted to reproduce the 5
velocities in the rotation curve. The weights used in the combination
have been derived from the relative intensities in the continuum of
the 5 one-dimensional galaxy spectra used to obtain the rotation
curve. The comparison of the velocity dispersion in the simulated
spectrum with the one used to broaden the stellar template spectra
provides an estimate of the amount of broadening due to the rotation.

Given the accuracy with which radial velocities can be measured, and
the small number of points used to define the rotation curve, rotation
velocities smaller than $\simeq$20 \kms\ cannot be reliably
measured. Within the sample presented here, 75 galaxies do not show
detectable rotation. The remaining 137 have a median rotation velocity
of 40 \kms, with the largest measured velocity being 130 \kms. Note
that we had no a priori information on the position angle of the axis
of rotation of each galaxy. This rotation is responsible for an
average broadening of the LOSVD of $\simeq$4\%, with a maximum of
$\simeq$9\% for the largest rotation velocities.

\subsection{Aperture Correction}

Because of radial gradients in the stellar velocity dispersion within
early-type galaxies, the velocity dispersion measured in a given
galaxy depends on the size of the aperture that is used to extract its
spectrum.  When spectra are obtained with a fixed aperture size (in
arcsec), this aperture covers regions of different physical size (in
kiloparsec) on galaxies at different distances, and this produces
biased velocity dispersion measurements. Therefore fixed aperture
measurements must be corrected to a standard, distance independent,
aperture size. Here all measurements are corrected to the value one
would obtain for a galaxy at the distance of the Coma cluster, using
the relation derived by J{\o}rgensen, Franx
\& Kj{\ae}rgaard (1995\markcite{Jor95})
\begin{equation} \label{eq:ap_corr}
\log {{\sigma_{ap}}\over{\sigma_{norm}}} = -0.04~\log {{r_{ap}}\over{r_{norm}}}
= -0.04~\log {{d}\over{d_{Coma}}}
\end{equation}
where $\sigma_{ap}$ and $\sigma_{norm}$ are the measured and corrected
velocity dispersion, $r_{ap}$ and $r_{norm}$ are the projected angular
size of a given length at the distance of the observed galaxy and at
the reference distance used for the normalization, and $d$ and
$d_{Coma}$ are the distance of the cluster to which the given galaxy
belongs, and that of the Coma cluster.  Distances are derived from the
cluster redshifts, in the CMB reference frame.

\section{The spectroscopic sample}

Spectroscopic data for early-type galaxies in 8 clusters and rich 
groups of galaxies are presented here. These are the NGC 383 group,
NGC 507 group, A262, Cancer, A1367, Coma (A1656), Pegasus, A2634. A short
discussion on their properties, and on the membership criteria we used
to assign galaxies to these clusters, is presented in Paper III.  Our
spectroscopic sample is composed of two partly overlapping datasets:
our own observations, discussed in the previous sections, for a total
of 212 galaxies, and a compilation of data from the literature, which
includes 152 galaxies. For the Coma cluster, most of the data come
from the literature, while for the other clusters most of the data are
new. In all cases we have tried to complement the existing
observations in order to obtain approximately flux limited samples,
while having enough overlap with those observations, as to compare the
different velocity dispersion scales.

\subsection{New observations}

The criteria adopted to select galaxies for our spectroscopic
observations were as follows: CGCG magnitudes (Zwicky \etal\
1963-1968\markcite{CGCG}) were used for the sample selection, if
available. Otherwise our own eye estimates, calibrated on the same
scale, were used.  All galaxies with morphological type E, S0, and S0a
were included in the sample. A few objects, that were originally
classified as S0 or S0a, were later classified Sa on the basis of the
CCD imaging (see Paper III), but they were left in the sample if a
spectrum had already been taken.  Only galaxies with previously
available redshift measurements, so that their cluster membership was
already known, were targeted for observations.

No particular criterion was used to select the objects that would
provide an overlap with already available datasets. An effort was
made, however, to cover as large an interval of velocity dispersions
as possible.

Table 2 presents the relevant parameters for the 212 objects
for which we have obtained spectroscopic observations. Only
the first page is presented herefor guidance regarding its form 
and content . The table is available in its entirety in digital form
on the Journal Electronic Edition, or from the authors. 
The table is organized as follows: \newline
Col. 1: Galaxy name: if the galaxy is listed in the UGC catalog (Nilson 
1973\markcite{UGC}), the UGC number is given; if not, our internal coding 
number, from the private galaxy catalog of Giovanelli \& Haynes referred 
to as the AGC, is given. \newline
Cols. 2 and 3: the NGC or IC number, if available, and the CGCG field and 
ordinal number within that field, if the galaxy is listed in the CGCG 
(Zwicky \etal\ 1963-1968\markcite{CGCG}). For galaxies in the Coma cluster 
that are not listed in either one of those catalogs, the Dressler number 
(Dressler 1980\markcite{Dre80}) is given, if available. \newline
Cols. 4 and 5: Right Ascension and Declination, in 1950.0 epoch. \newline
Col. 6: morphological type code, in the RC3 (de Vaucouleurs \etal\ 
1991\markcite{RC3}) scheme (-5 for E, -2 for S0, 0 for S0a). Morphological 
types were derived from joint inspection of the blue plates of the Palomar 
Observatory Sky Survey (POSS), and of the CCD images discussed in Paper 
III. \newline
Col. 7: Measured heliocentric velocity, in \kms. \newline
Col. 8: Measured stellar velocity dispersion, in \kms. \newline
Col. 9: Measured rotation velocity, in \kms. If no data is present,
the spectrum had to low a S/N ratio to perform the measurement. \newline
Col. 10: Velocity dispersion, in \kms, corrected for galaxy rotation. \newline
Col. 11: Logarithm of the rotation-corrected velocity dispersion. \newline
Col. 12: Uncertainty on the logarithm of the velocity dispersion. \newline
Col. 13: Aperture-corrected logarithm of the velocity dispersion. \newline
Col. 14: A reference code indicates if other velocity dispersion measurements
were available in the literature, and their source. If an asterisk appears 
in this column, special comments on the object are included at the foot of 
the table.

A subset of these data has already been published in Paper I, 
limited to galaxies in the Coma and A2634 clusters. Since then, a few
galaxies have been reobserved, and the data combined to obtain new
measurements of the relevant photometric parameters. The data presented
here therefore supersede that previous report. 

\subsection{Data from the literature}

Velocity dispersion measurements are available in the literature for
each of the clusters in our sample. The number is large only in the
case of the Coma cluster, where measurements from Dressler
(1987\markcite{Dre87}), Davies \etal (1987\markcite{Dav87}), Faber
\etal (1989\markcite{Faber89}), and Lucey \etal (1991\markcite{Lucey91}) 
are available for a total of 103 galaxies, and in the case of A2634,
where measurements from Lucey \etal (1997\markcite{Lucey97}) are
available for 37 galaxies. For the remaining clusters, we have found
data in Faber \etal (1989\markcite{Faber89}) for 6 galaxies in A262, 7
galaxies in A1367, 1 galaxy in Cancer, 4 galaxies in Pegaus, 12
galaxies in the NGC 507 and NGC 383 groups.  For the brightest galaxy
in some of the clusters multiple measurements from various sources are
available (see the footnotes to Table 2).  In all clusters, we have
limited our search to galaxies for which we have obtained I band
photometric measurements (see Paper III).  Therefore a larger
spectroscopic dataset than the one reported here might be available in
some cases.  Table 3 lists the relevant parameters for the 170 objects
for which data are derived from the literature. Only the first page is
presented here; the entire table is available in digital form on the
Journal Electronic Edition, or from the authors.  The table is
organized as follows: \newline 
Cols. 1 to 6: same as in Table 2. \newline 
Col. 7: Heliocentric velocity, in \kms. \newline 
Col. 8: Stellar velocity dispersion, in \kms. \newline 
Col. 9: Logarithm of the velocity dispersion. \newline 
Col. 10: Uncertainty on the logarithm of the velocity dispersion.  
A common value for each main dataset was adopted, corresponding to the 
typical uncertainty quoted by the authors. \newline 
Col. 11: Aperture-corrected logarithm of the velocity dispersion. 
\newline Col. 12: Reference code.

\section{Internal and external comparison}

Repeated observations of a fraction of the galaxies in our sample
provide the opportunity to evaluate the internal consistency of the
spectroscopic measurements. However, the number of repeated
observations is rather small, because spectroscopic observations are
quite expensive in terms of telescope time. In total, 27 pairs of
observations are available for an internal comparison. Table 4 lists
the results of the comparison of radial velocity and velocity
dispersion. Mean statistical uncertainties associated with the
derivation of those parameters are included in the Table, multiplied
by a $\sqrt{2}$ factor to allow a direct comparison with the rms
scatter in the observations (assuming equal error contributions to the
rms scatter).  The rms scatter in the radial velocity comparison (see
Figure 2a) is surprisingly large (29 \kms), but it is mostly due to
just 2 galaxies.  After removing these extreme cases the scatter is
reduced to 19 \kms, in agreement with the statistical uncertainties on
the radial velocity determinations. Unfortunately with such small
number statistics it is difficult to decide whether these 2 objects
are truly outliers. However we point out that in bot cases one of the
two spectra being compared has an extremely low signal-to-noise ratio,
and therefore the uncertainty on the determination of the
corresponding spectroscopic parameters is expected to be large.

The comparison of the velocity dispersions is presented in Figure 2b,
and shows that, as expected, the accuracy of the dispersion
determinations for very low velocity dispersion objects 
($\sigma < 100$\kms) is lower than that for higher velocity dispersion
objects. The rms scatter measured for the 6 low velocity dispersion
objects is in fact more than twice as large as the one measured for
the remaining 21 objects. When these low velocity dispersion objects
are removed from the sample, we find a fair agreement between the
expected uncertainties and the observed dispersion in the
measurements.

The absolute consistency of the velocity dispersion measurements is
evaluated through a comparison with measurements taken from the
literature. There are 62 observations for 57 galaxies for which this
comparison is possible. The results of such a comparison are presented
in Table 5 and in Figure 3, where the difference (us - literature) is
plotted vs. the mean value of the velocity dispersion, derived
combining the two measurements. Five extremely discrepant objects are
removed from this comparison. For two objects the measurement we have
obtained for their velocity dispersion is affected by a large
uncertainty. For the remaining three no easy explanation for the
observed discrepancy is available. It is clear that there is good
agreement with the observations of Dressler (1987\markcite{Dre87}),
Davies \etal (1987\markcite{Dav87}), and Faber \etal
(1989\markcite{Faber89}), collectively indicated as 7S in Table 5, and
also with the observations of Lucey \etal (1997\markcite{Lucey97}).
No significant offset is present, and the observed rms scatter of the
differences in the values of $\log \sigma$ is in good agreement with
an expected one, based on the 6\% median uncertainty of our
measurements, and on an average 10\% uncertainty for the measurements
in the literature.  A small discrepancy is present when the Coma
cluster observations of Lucey \etal (1991\markcite{Lucey91}) are
considered, but the sample is probably too small to derive robust
conclusions. We notice that the comparison between Lucey \etal
(1991\markcite{Lucey91}) and Faber \etal (1989\markcite{Faber89})
datasets, based on a larger sample of galaxies, shows good relative
agreement (see Lucey \etal 1991\markcite{Lucey91}). Given the good
agreement between the different datasets, all measurements presented
in Table 2 and 3 will be merged for the FP work that will be discussed
in Scodeggio \etal (1998b\markcite{pap5}).

\acknowledgments
We would like to thank John Henning, Mike Doyle, Dave Tennant, and Bob
Thicksten for their technical support during the various observing
runs at Palomar, Juan Carrasco for his masterful handling of many
observing nights, and Gary Wegner for helpful discussions on the
comparison between different spectroscopic datasets.  This research
has made use of the NASA/IPAC Extragalactic Database (NED) which is
operated by the Jet Propulsion Laboratory, California Institute of
Technology, under contract with the National Aereonautics and Space
Administration.  This research is part of the Ph.D. Thesis of MS, and
is supported by the NSF grants AST94--20505 and AST96--17069 to RG and
AST95--28860 to MPH.

\begin{table}
\tablenum{1}
\caption[]{Spectroscopic observing runs}
{\small
\begin{tabular}{lcc}
\\
\tableline
\tableline
\\
                 &  Runs 1--6             &  Runs 7--8      \\
\\
\tableline   
\\       
Telescope           & Palomar 5m         & Palomar 5m      \\
Instrument          & DS-red camera      & DS-red camera   \\
CCD                 & TI 305 \#8         & Tek1024         \\
Pixels              & 800 x 300          & 1024 x 300      \\
Read-out noise      & 11 e$^-$           & 7.5 e$^-$       \\
Gain                & 1.7 e$^-$/DU       & 2.0 e$^-$/DU    \\
Spatial scale       & 0.61 ''/pixel      &  0.46 ''/pixel  \\
Slit aperture       & 2 arcsec           & 2 arcsec        \\
Grating             & 1200 l/mm          & 1200 l/mm       \\
Dispersion          & 0.82 \AA/pixel     & 0.66 \AA/pixel  \\
Spectr. resol.      & 2.2 \AA\ (129 km/s) & 2.2 \AA\ (129 km/s)  \\
Wavelength cover.   & 4960-5620 \AA      & 4960-5660 \AA   \\
\\
\end{tabular}
\begin{tabular}{lcccc}
\tableline
\tableline
\\
                 &  Run 1     &  Run 2           & Run 3   &  Run 4    \\
\\
\tableline    
\\       
Dates        & Sep. 23-26 1992  & Sep. 17-20 1993  & Mar. 3-6 1994 & Apr. 28-29 1994\\
Nights used  & 2                & 4                & 3             & 1    \\
Nr. of galaxies & 16            & 64               & 31            & 6    \\
\\
\tableline
\tableline
\\
             &  Run 5            & Run 6    &  Run 7     &  Run 8     \\
\\
\tableline    \\ 
Dates        & Sep. 6-12  1994  & Aug. 23-25 1995  & Mar. 19-20 1996 & Sep. 20 1996  \\ 
Nights used      & 7                 & 3           & 1                  & 1   \\
Nr. of galaxies  & 72                & 28          & 5                  & 8  \\
\\
\tableline
\tableline
\end{tabular}
}

\end{table}

\begin{deluxetable}{lrrrr}
\tablewidth{0pt}
\tablenum{4}
\tablecaption{Spectroscopic data quality: internal comparison}
\tablehead{
\colhead{Parameter} & \colhead{N$_{gal}$} & \colhead{mean} &
\colhead{rms} & \colhead{mean}  \nl
&& \colhead{difference} & \colhead{scatter} & \colhead{uncertainty}
}
\startdata
$V_{hel}$ (all)   & 27  &   5.8\phm{aa}  &  28.9\phm{aa}  &   18.0\phm{aa} \nl
$V_{hel}$ (2 most & 25  &  -2.2\phm{aa}  &  19.3\phm{aa}  &   18.0\phm{aa} \nl
discr. excluded)   \nl

$\log\sigma$ (all)              & 27 &  0.002\phm{aa} &  0.055\phm{aa}  &  0.031\phm{aa}  \nl
$\log\sigma ~(\sigma>100$ \kms) & 21 &  0.002\phm{aa} &  0.041\phm{aa}  &  0.037\phm{aa} \nl
$\log\sigma ~(\sigma<100$ \kms) &  6 &  0.001\phm{aa} &  0.096\phm{aa}  &  0.037\phm{aa} \nl
\enddata
\end{deluxetable}

\begin{deluxetable}{lrrr}
\tablewidth{0pt}
\tablenum{5}
\tablecaption{Spectroscopic data quality: external comparison}
\tablehead{
\colhead{Parameter} & \colhead{N$_{gal}$} & \colhead{mean} &
\colhead{rms}  \nl
&& \colhead{difference} & \colhead{scatter}
}
\startdata
$\log\sigma$ (all) & 56 & -0.004\phm{aa}  &  0.053\phm{a} \nl
$\log\sigma$ (7S)  &   29 &  -0.004\phm{aa}  &  0.050\phm{a} \nl
$\log\sigma$ (L97)  &  21  &  -0.001\phm{aa}   &   0.057\phm{a} \nl
$\log\sigma$ (L91)  &  6  &  -0.015\phm{aa}   &   0.056\phm{a} \nl
\enddata
\end{deluxetable}

\newpage

\plotone{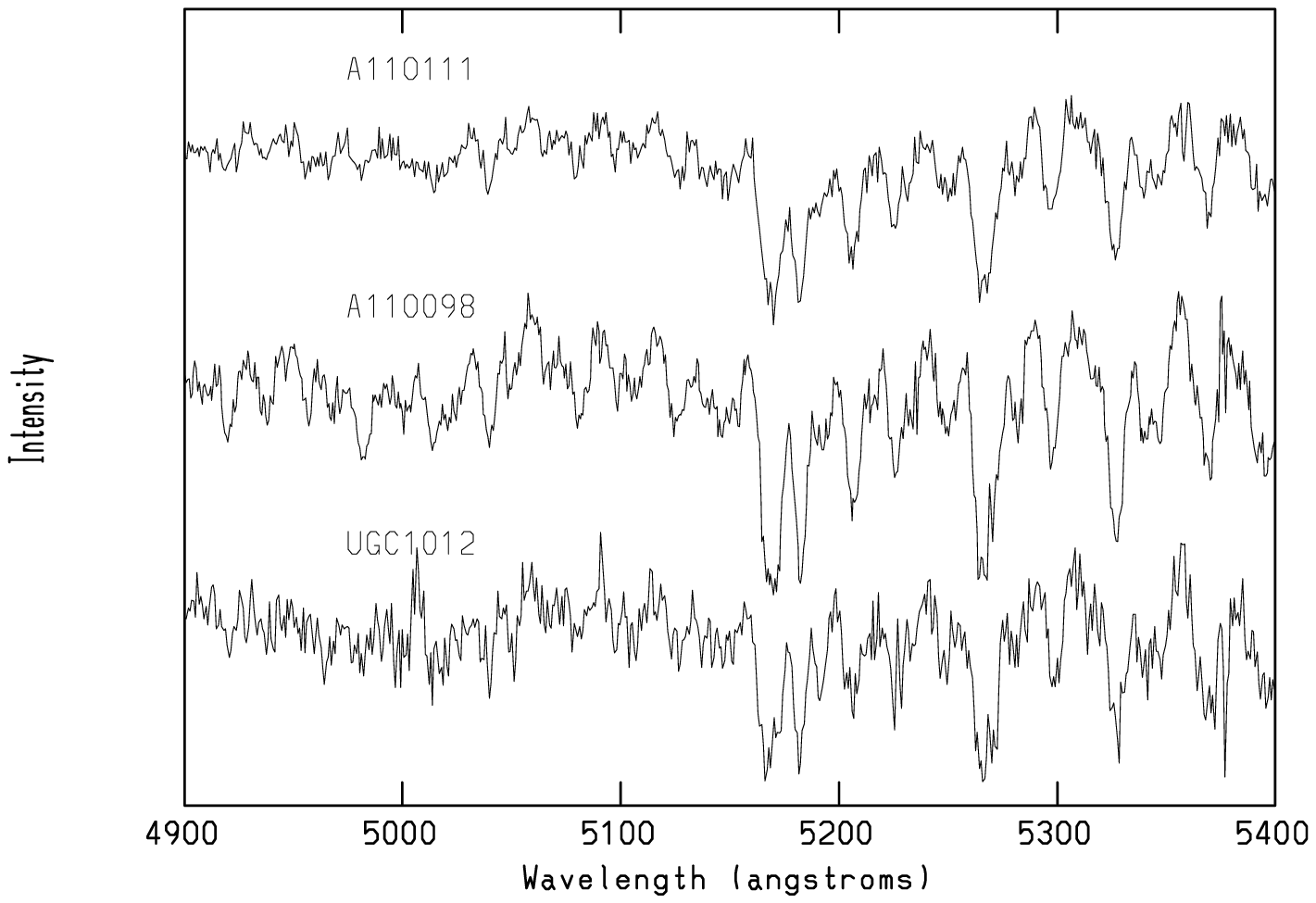}
\figcaption[fp_univers2_fig1a.ps]{Spectra of 6 different galaxies in our 
data-set.  These are divided into two groups, and all three glaxies in
one group have very similar velocity dispersion (approximately 140 and
230 \kms, respectively), but their spectra have different S/N
ratio. These spectra were chosen to provide an example of a spectrum
at the upper quartile, median, and lower quartile (top to bottom) in
the distribution of S/N values for the spectra in our data-set. }

\setcounter{figure}{0}
\plotone{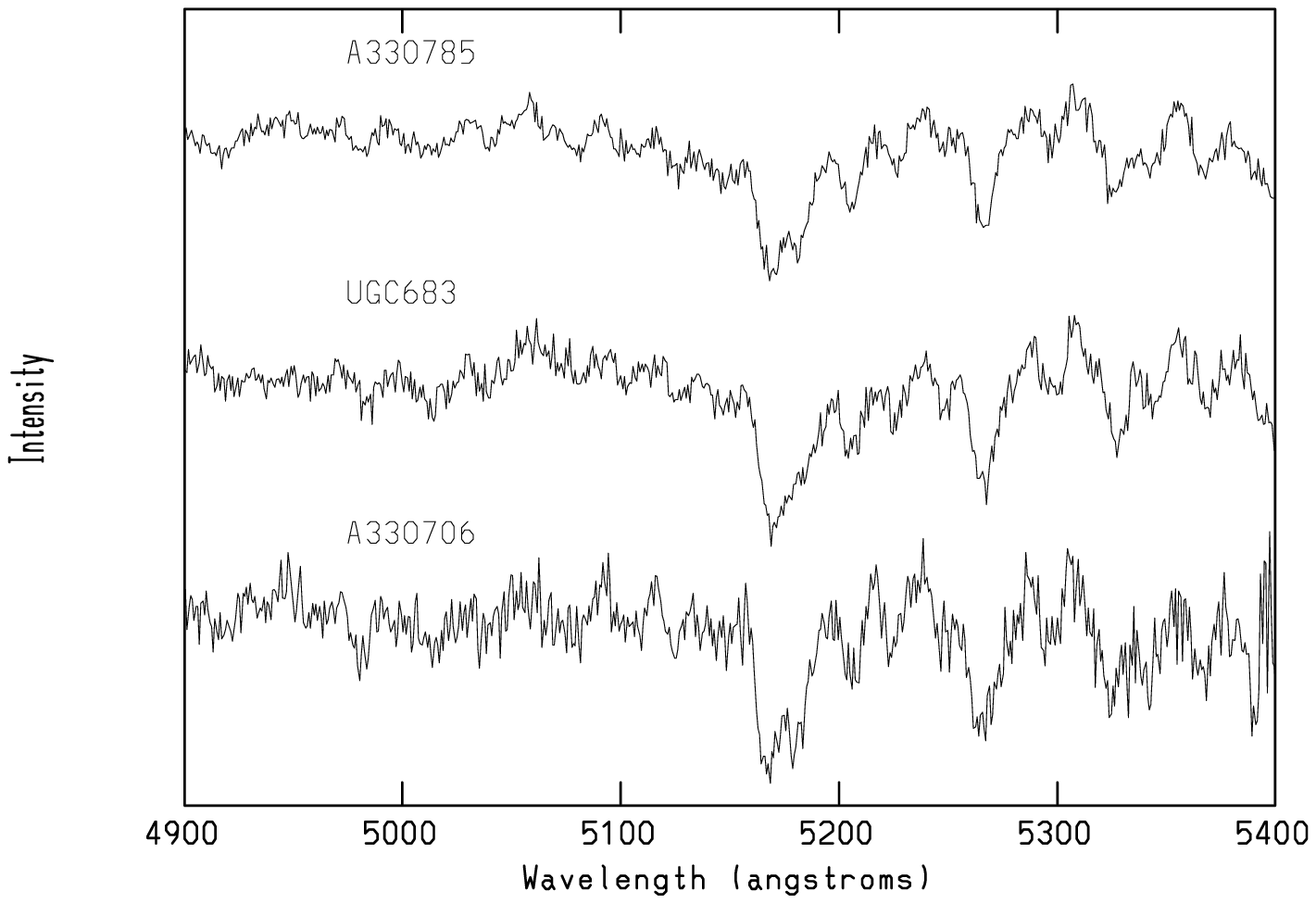}
\figcaption[fp_univers2_fig1b.ps]{(Continued)}

\plotone{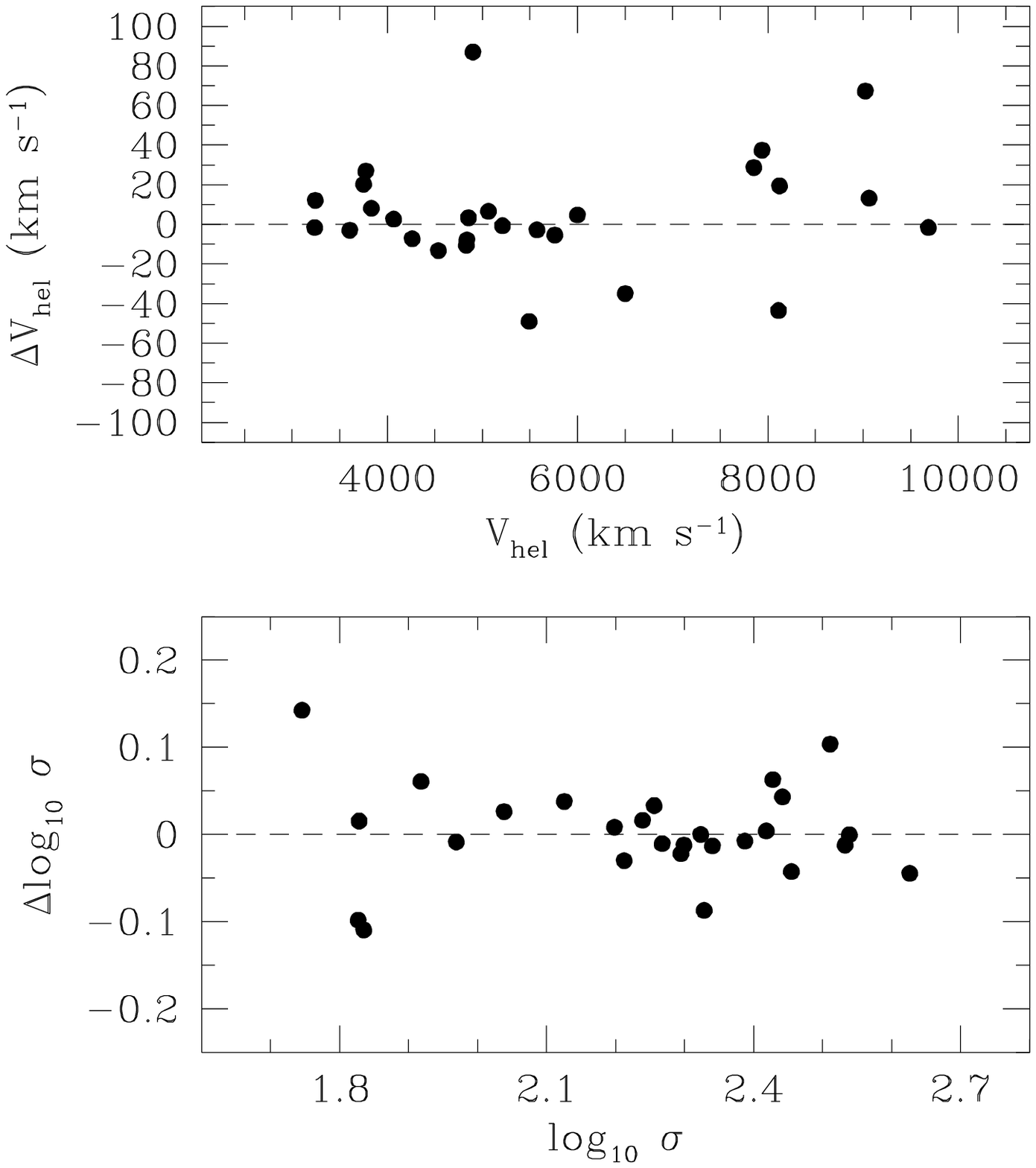}
\figcaption[fp_univers2_fig2.ps]{Internal comparison of the newly derived 
spectroscopic parameters. $\Delta V_{hel}$ and $\Delta log_{10}
\sigma$ are the differences between the values of the heliocentric
radial velocity and of the logarithm of the velocity dispersion
obtained from repeated observations of 27 galaxies.  They are plotted
as a function of the mean value of the relative parameter, obtained
combining the different measurements.}

\plotone{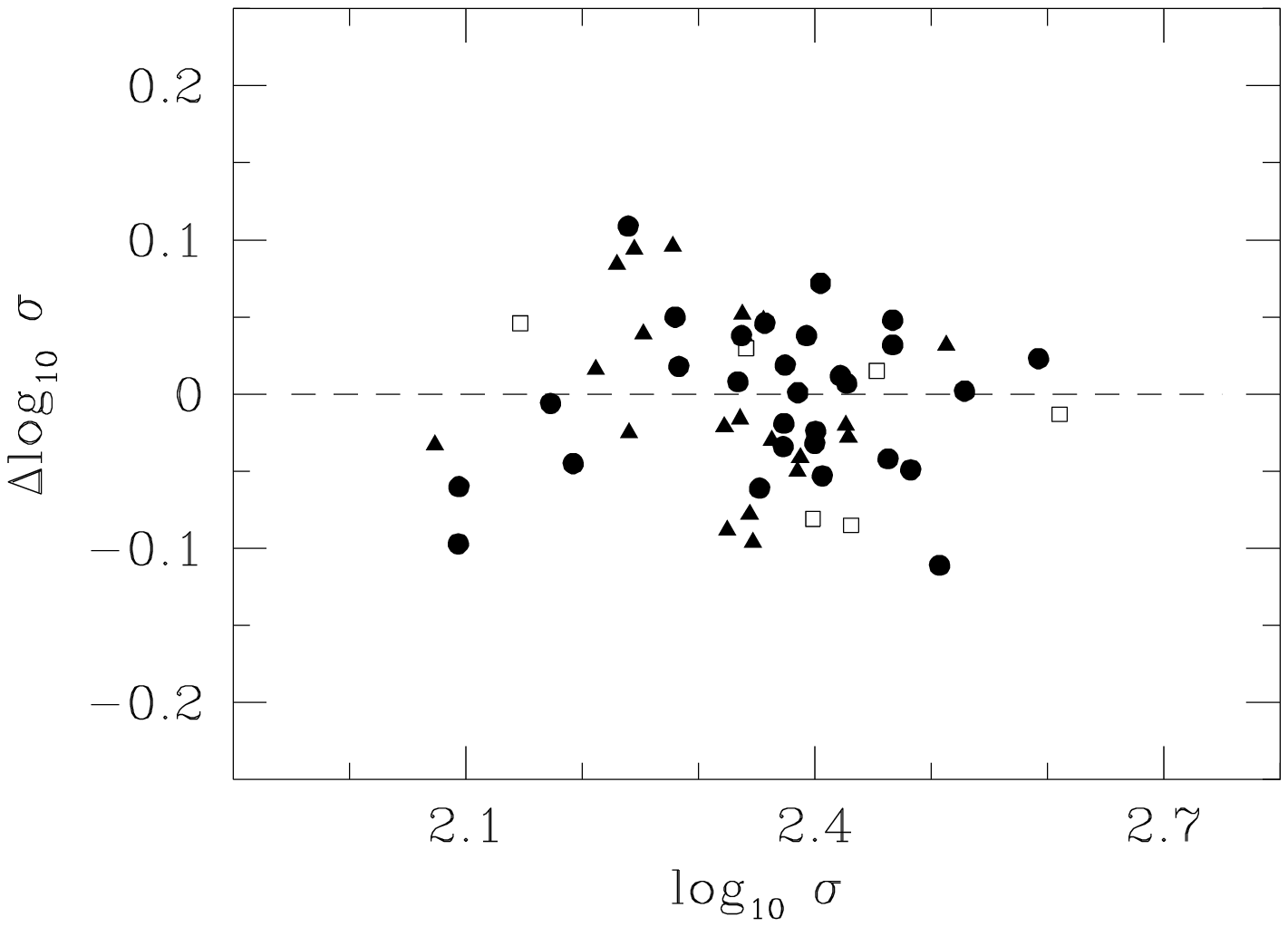}
\figcaption[fp_univers2_fig3.ps]{External comparison of the velocity 
dispersion measurements for galaxies in our sample. $\Delta log_{10}
\sigma$ is the difference between the value of the logarithm of the
velocity dispersion we obtained for a given galaxy, and the value
found in the literature. This difference is plotted against mean value
of this parameter, obtained combining the different
measurements. Filled circles identify the comparison with the
measurements of Faber \etal (1989), filled triangles the comparison
with the measurements of Lucey \etal (1997), while open squares
identify the comparison with the data of Lucey \etal (1991).}


\begin{references}
\reference{Ben90} Bender, R. 1990, \aap, 229, 441 
\reference{Dav83} Davies, R.L., \& Illingworth, G. 1983, \apj, 266, 516
\reference{Dav87} Davies, R.L., Burstein, D., Dressler, A., Faber, S.M., 
Lynden-Bell, D., Terlevich, R.J., \& Wegner, G. 1987, \apjs, 64, 581 
\reference{RC3} de Vaucouleurs, G., de Vaucouleurs, A., Corwin, H.G., Buta, R., 
Paturel, G., \& Fouqu\'e, P. 1991, ``Third Reference Catalogue of Bright 
Galaxies'' (New York: Springer) (RC3) 
\reference{DD} Djorgovski, S.G., \& Davis, M. 1987, \apj, 313, 59 
\reference{Djorg88} Djorgovski, S.G., De Carvalho, R.R., \& Han, M.S. 1988, in ``The 
Extragalactic Distance Scale'', ed. S. van den Bergh \& J. Pritchet, p. 329 
\reference{Dre80} Dressler, A. 1980, \apjs, 42, 565 
\reference{Dre87} Dressler, A. 1987, \apj, 317, 1 
\reference{7S} Dressler, A., Lynden-Bell, D., Burstein, D., Davies, R.L., Faber,
S.M., Terlevich, R.J., \& Wegner, G. 1987, \apj, 313, 42 
\reference{FJ} Faber, S. M., \& Jackson, R. E. 1976, \apj, 204, 668
\reference{Faber89} Faber, S.M., Wegner, G., Burstein, D., Davies, R.L., Dressler, A.,
Lynden-Bell, D., \& Terlevich, R.J. 1989, \apjs, 71, 173 
\reference{Fish95} Fisher, D., Illingworth, G., \& Franx, M. 1995, \apj, 438, 539
\reference{Franx89} Franx, M., Illingworth, G., \& Hechman, T. 1989, \apj, 344, 613 
\reference{G97a} Giovanelli, R., Haynes, M.P., Herter, T., Vogt, N.P., da Costa, 
L.N., Freudling, W., Salzer, J.J. and Wegner, G. 1997a, \aj, 113, 22 (G97a) 
\reference{G97b} Giovanelli, R., Haynes, M.P., Herter, T., Vogt, N.P., da Costa, 
L.N., Freudling, W., Salzer, J.J. and Wegner, G. 1997b, \aj, 113, 53 (G97b) 
\reference{Guz92} Guzm{\'a}n, R., Lucey, J.R., Carter, D., \& Terlevich, R.J. 1992, 
\mnras, 257, 187 
\reference{Jor95} J{\o}rgensen, I., Franx, M., \& Kj{\ae}rgaard, P. 1995, 
\mnras, 276, 1341 
\reference{Jor96} J{\o}rgensen, I., Franx, M., \& Kj{\ae}rgaard, P. 1996, 
\mnras, 280, 167 
\reference{Lucey91} Lucey, J.R., Guzm{\'a}n, R., Carter, D., \& Terlevich, R.J. 
1991, \mnras, 253, 584 
\reference{Lucey97} Lucey, J.R., Guzm{\'a}n, R., Steel, J., \& Carter, D., 1997, 
\mnras, 287, 899
\reference{Mal81} Malumuth, E.M., \& Kirshner, R.P. 1981, \apj, 251, 508
\reference{UGC} Nilson, P. 1973, ``Uppsala General Catalogue of Galaxies'' 
(Uppsala Astron. Obs. Ann., Vol. 6) (UGC) 
\reference{Oeg91} Oegerle, W.R., \& Hoessel, J.G. 1991, \apj, 375, 15
\reference{dbsp} Oke, J.B., \& Gunn, J.E. 1982, \pasp, 94, 586 
\reference{SSBS} Sargent, W.L.W., Schechter, P.L., Boksenberg, A., \& Shortridge, K.,
1977, \apj, 212, 326 
\reference{Schro96} Schroeder, A., 1996, PhD Thesis, Univ. of Basel 
\reference{pap1} Scodeggio, M., Giovanelli, R., \& Haynes, M.P. 1997a,
\aj, 113, 101 (Paper I) 
\reference{pap2} Scodeggio, M., Giovanelli, R., \& Haynes, M.P. 1997b,
\aj, 113, 2087 (Paper II) 
\reference{pap3} Scodeggio, M., Giovanelli, R., \& Haynes, M.P. 1998a,
submitted to \aj (Paper III) 
\reference{pap5} Scodeggio, M., Giovanelli, R., \& Haynes, M.P. 1998b,
in preparation 
\reference{Ton84} Tonry, J.L. 1984, \apj, 279, 13
\reference{Ton85} Tonry, J.L. 1985, \aj, 90, 2431
\reference{TD} Tonry, J.L., \& Davis, M. 1979, \aj, 84, 1511 
\reference{Ton81} Tonry, J.L., \& Davis, M. 1981, \apj, 246, 666
\reference{TF} Tully, R.B., \& Fisher, J.R. 1977, \aap, 54, 661 
\reference{CGCG} Zwicky, F., Herzog, E., Wild, P., Karpowicz, M., \& Kowal, C. 
1961--1968, ``Catalogue of Galaxies and Clusters of Galaxies'' 
(Pasadena: California Institute of Technology) 

\end{references}
\end{document}